\documentclass[%
 aip,
 jmp,%
 amsmath,amssymb,
 reprint,%
 twocolumn,
 superscriptaddress,
]{revtex4}

\usepackage{stmaryrd}
\usepackage{graphicx}
\usepackage{dcolumn}
\usepackage{bm}
\usepackage{textcomp}
\usepackage{multirow,amssymb,amsbsy,amsmath}
\usepackage{ctable}
\usepackage{colortbl}
\usepackage{amsmath}

\begin{document}

\title{High-efficiency spontaneous terahertz emission from electrically-excited plasmons in AlGaN/GaN two-dimensional electron gas}

\author{Hua Qin}\email[Corresponding author, ]{hqin2007@sinano.ac.cn}
\affiliation{
Key Laboratory of Nanodevices and Applications, 
Suzhou Institute of Nano-tech and Nano-bionics (SINANO), 
Chinese Academy of Sciences, 
398 Ruoshui Road, Suzhou 215123, P.~R.~China}

\author{Yao Yu}
\affiliation{
Key Laboratory of Nanodevices and Applications, 
Suzhou Institute of Nano-tech and Nano-bionics (SINANO), 
Chinese Academy of Sciences, 
398 Ruoshui Road, Suzhou 215123, P.~R.~China}
\affiliation{
Graduate University of Chinese Academy of Sciences, Beijing 100049, P.~R.~China}

\author{Jiandong Sun}\email[Corresponding author, ]{jdsun2008@sinano.ac.cn}
\affiliation{
Key Laboratory of Nanodevices and Applications, 
Suzhou Institute of Nano-tech and Nano-bionics (SINANO), 
Chinese Academy of Sciences, 
398 Ruoshui Road, Suzhou 215123, P.~R.~China}

\author{Zhongxin Zheng}
\affiliation{
Key Laboratory of Nanodevices and Applications, 
Suzhou Institute of Nano-tech and Nano-bionics (SINANO), 
Chinese Academy of Sciences, 
398 Ruoshui Road, Suzhou 215123, P.~R.~China}
\affiliation{
Graduate University of Chinese Academy of Sciences, Beijing 100049, P.~R.~China}

\author{Yongdan Huang}
\altaffiliation[Now at:]{
Key Laboratory of Infrared Imaging Materials and Detectors, 
Shanghai Institute of Technical Physics, Chinese Academy of Sciences, Shanghai 200083, P.~R.~China}
\author{Xingxin Li}
\author{Yu Zhou}
\author{Dongmin Wu}
\author{Zhipeng Zhang}
\author{Cunhong Zeng}
\author{Yong Cai}
\author{Xiaoyu Zhang}
\author{Baoshun Zhang}
\affiliation{
Key Laboratory of Nanodevices and Applications, 
Suzhou Institute of Nano-tech and Nano-bionics (SINANO), 
Chinese Academy of Sciences, 
398 Ruoshui Road, Suzhou 215123, P.~R.~China}

\author{Xuecou Tu}
\author{Gaochao Zhou}
\author{Biaobing Jin}
\author{Lin Kang}
\author{Jian Chen}
\author{Peiheng Wu}
\affiliation{
Research Institute of Superconductor Electronics (RISE), 
School of Electronic Science and Engineering, 
Nanjing University, Nanjing 210093, P. R. China.}

\date{\today}  

\begin{abstract}
The advance of terahertz science and technology yet lays wait for 
the breakthrough in high-efficiency and high-power solid-state terahertz sources applicable at room temperature. 
Plasmon in two-dimensional electron gas (2DEG) has long been pursued as a type of promising active medium for terahertz emitters. 
However, a high wall-plug efficiency sufficient for high-power operation has not been achieved.  
Here we report spontaneous terahertz emission tunable in a frequency range from 0.3 to 2.5~THz 
through electrical excitation of plasmon or plasmon-polariton modes in a grating-gate coupled AlGaN/GaN 2DEG at 7~K. 
A wall-plug efficiency ranging from $10^{-6}$ to $10^{-1}$ is found by tuning the energy (4.5~eV to 0.5~eV) of tunneling electrons 
between the grating gates and the 2DEG. 
The total emission power, albeit below 15~nW, is already detectable by using a direct terahertz field-effect transistor detector 
based on the same type of 2DEG. 
The findings indicate that both high-efficiency and high-power terahertz plasmon-polariton emission 
could be obtained by a proper injection of tunneling electrons with low energy and high current density.
\end{abstract}

\maketitle

Efficient solid-state terahertz emitters/oscillators are one of the most important devices for terahertz applications~\cite{miles,tonouchi,chattopadhyay}. 
However, both electronic and photonic devices become less efficient in generating electromagnetic wave 
in the frequency range centered around 1~THz especially at room temperature. 
Great efforts are being made in searching of more efficient terahertz emitters. 
Among them plasmon-based terahertz emitters can be viewed as a marriage of the electronic approach and the photonic approach. 
Plasmons in a high-electron-mobility two-dimensional electron gas (2DEG) typically formed in semiconductor heterostructures such as AlGaAs/GaAs and AlGaN/GaN, 
two-dimensional materials such as graphene, can be continuously tuned from a few 100 GHz to a few THz by varying the electron density~\cite{gornik,allen,tsui,hopfel,hirakawa,maranowski,bakshi99}. 
As collective charge oscillations in the terahertz frequency range, 
plasmon in semiconductors has long been pursued for solid-state terahertz emitters. 
An electrically-driven terahertz plasmon emitter embodies two core processes, 
namely, the electrical excitation of plasmons and the terahertz emission from plasmons. 
Unfortunately, electrical excitation of solid-state plasma wave is extremely inefficient due to the high damping rate of plasmons~\cite{sst-plasma}. 
Although there exists efficient grating coupler, antenna, cavity and other similar means,  
terahertz plasmon emission is also strongly limited by the short life time of solid-state plasma wave~\cite{mackens,alsmeier,wilkinson,popov,otsuji}.  
When the above two inefficient processes are cascaded in an emitter, no high-power emitter could be made. 
Instabilities of driven plasma in nano-structured or ballistic 2DEG channels 
has been proposed and studied to improve the efficiency in electrical excitation of plasmons~\cite{cen,dyakonov,bakshi95,tralle,mikhailov,kempa99}. 
However, the required threshold drift velocity usually close to or larger than the electron's Fermi velocity 
is hardly achievable since electrons inevitably suffer from strong scattering, e.g., by longitudinal optical (LO) phonons. 
Consequently, 2DEG and its host crystal are strongly heated 
and blackbody-like emission from hot electrons and lattice surpasses the possible plasmonic terahertz emission~\cite{hirakawa93,bhatti,zhou,zheng}.  
Recently, Ryzii and Shur proposed self-excitation of plasmons in a 2DEG channel 
by extracting tunneling electrons from the channel to the field-effect gate ~\cite{ryzhii02}. 
Different in energy scale, this idea is in fact in close analogy to plasmon excitation in thin foil of aluminum 
by an electron beam with energy in keV range observed by Ruthemann and Lang in 1960's~\cite{ruthemann,lang}.
When the injection energy is tuned, excitation of unwanted quasi-particles, e.g., LO phonons could be suppressed. 
Furthermore, it is possible to strongly couple plasmon modes with the terahertz electromagnetic cavity modes so that 
plasmon-polariton modes could be constructed~\cite{satou,geiser-prl2012,geiser-apl2012,huang2013,huang2015}. 
As a hybrid state of a terahertz cavity mode (light) and a plasmon mode (matter), 
a plasmon-polariton mode allows for high-efficiency energy transfer from the plasmon mode to the terahertz cavity mode, and vice versa. 

In this letter, we report on tunable, threshold-less and spontaneous terahertz emission from plasmons and plasmon polaritons in a grating-gate coupled 2DEG 
without and with a terahertz Fabry-P\'erot cavity excited by tunneling electrons between the grating gates and the 2DEG, respectively. 
A conversion efficiency in an order of $10^{-1}$ is achieved. 
The results confirm that solid-state plasma wave can be used as an active terahertz medium and 
plasmon-based terahertz emitters with both high efficiency and high power 
could be realized by further improving the injection efficiency and the quality factors of the emission modes.

\begin{figure}[!htp]
 \includegraphics[width=.45\textwidth]{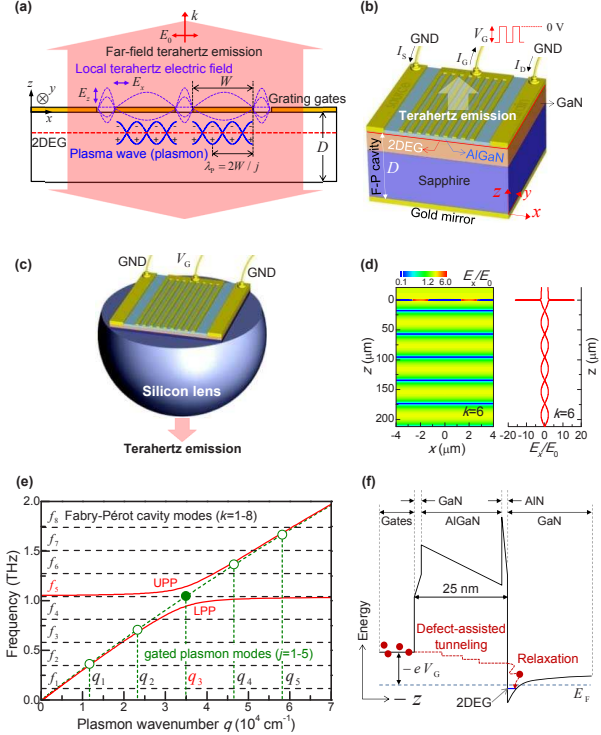}
\caption{
Schematics of the core terahertz plasmonic device structure (a), the cavity emitter (b) and the lens emitter (c). 
(d)~Terahertz field of the 6th F-P cavity mode.  
(e)~Dispersion relations for the F-P cavity modes ($k=1,2,\cdots 8$), the gated plasmon modes ($j=1,2,3,\cdots,5$) with a certain electron density 
and the plasmon polariton modes (LPP, UPP) from the strongly-coupled the 5th F-P mode and the 3rd plasmon mode. 
(f)~Conduction-band diagram of the 2DEG under a negative gate voltage. 
}\label{fig_1}
\end{figure}

Two types of emitters, namely the \textit{cavity} emitter and the \textit{lens} emitter, 
are constructed based on the same core plasmonic device but with distinctly different terahertz electromagnetic environment. 
The core plasmonic device is a grating-gate coupled 2DEG based on an AlGaN/GaN heterostructure grown on sapphire substrate, as shown in Fig.~\ref{fig_1}(a). 
The ungated 2DEG has an electron density of $n_0 \approx 1.2\times 10^{13}~\mathrm{cm^{-2}}$ and 
an electron mobility of $\mu \approx 1.5\times 10^4~\mathrm{cm^2/Vs}$ at 7~K. 
The grating gates biased at a certain negative voltage $V_\mathrm{G}$ tune continuously the electron density $n_\mathrm{s}$ 
and the pinch-off voltage at which the electron density under a gate is fully depleted is $V_\mathrm{T}\approx -4.2$~V. 
The grating has a pitch distance of $L=4.0~\mathrm{\mu m}$ (in direction $x$), 
a width of 4~mm (in direction $y$) and a thickness of 200~nm (in direction $z$). 
Each gate has a length of $W=2.9~\mathrm{\mu m}$ resulting ungated 2DEG strips with a length about $L-W=1.1~\mathrm{\mu m}$. 
The cavity emitter is made by mounting the device on a gold-plated chip carrier which acts as a gold mirror on the backside of sapphire substrate, 
as schematically shown in Fig.~\ref{fig_1}(b). 
Serving as the main body of a terahertz Fabry-P\'erot (F-P) cavity, 
the sapphire substrate is thinned to a thickness of $D=212~\mathrm{\mu m}$ so as to reduce the number of terahertz modes with frequency below 3.0~THz.
As a counterpart to the cavity emitter, 
the lens emitter is realized by mounting the same plasmonic device on the flat surface of a hyperspherical silicon lens, as schematically shown in Fig.~\ref{fig_1}(c).
The grating gate above the 2DEG channel offers three functions: 
(1), formation of plasmon cavities in 2DEG under each gate; 
(2), coupling of terahertz electromagnetic wave and plasmons;
(3), injection/extraction of tunneling electrons into/from the 2DEG. 

The grating gates form about 1000 individual plasmon cavities in the 2DEG with the well-known dispersion relation~\cite{allen} 
$\omega_{\mathrm{P}j}=\sqrt{{n_\mathrm{s}e^2}q_\mathrm{j}/{2m^*\epsilon_0\bar\epsilon}}$, 
where $\omega_{\mathrm{P}j}=2\pi f_{\mathrm{P}j}$ is the angular frequencies, 
$e$ is the elementary charge, $j=1, 2, 3, \cdots$ is the plasmon mode index, $q_j=j\pi/W$ is the plasmon wave vector, $m^*$ is the effective electron mass, 
$\epsilon_\mathrm{0}$ is the vacuum permittivity, and $\bar\epsilon$ is the effective permittivity of the 2DEG with the grating coupler.
The charge-density oscillations and the electric fields corresponding to plasmon mode $j=3$ 
are schematically shown in Fig.~\ref{fig_1}(a). 
Plasmon modes excited in the lens emitter are ready to emit terahertz electromagnetic wave due to the grating gates. 
In the cavity emitter，a plasmon mode becomes emissive 
only when it is tuned to be in resonance with one of the F-P cavity modes, i.e., $\omega_{\mathrm{P}j}=\omega_k$. 
Such F-P modes correspond to standing waves with transverse electric (TE) field component $E_x$, 
a wave node at the mirror side ($z=-D$) and an anti-node at the grating side ($z=0$).  
The mode frequency can be expressed as $\omega_k = (2k-1){\pi c}/{2\bar nD}$，
where $\omega_k=2\pi f_k$ is the angular mode frequency with mode index of $k=1, 2, 3, \cdots$, 
$c$ is the speed of light in vacuum, and $\bar n \approx 3.0 $ is the effective refractive index of the sapphire cavity.  
The field distribution of F-P mode with $k=6$ is simulated and presented in Fig.~\ref{fig_1}(d). 
Strong field enhancement occurs near the edges of gates and 
allows for strong coupling between the plasmon modes and the F-P modes. 
Formation of a lower and an upper plasmon-polariton state (LPP, UPP) is illustrated in Fig.~\ref{fig_1}(e). 

\begin{figure}[!htp]
\includegraphics[width=.45\textwidth]{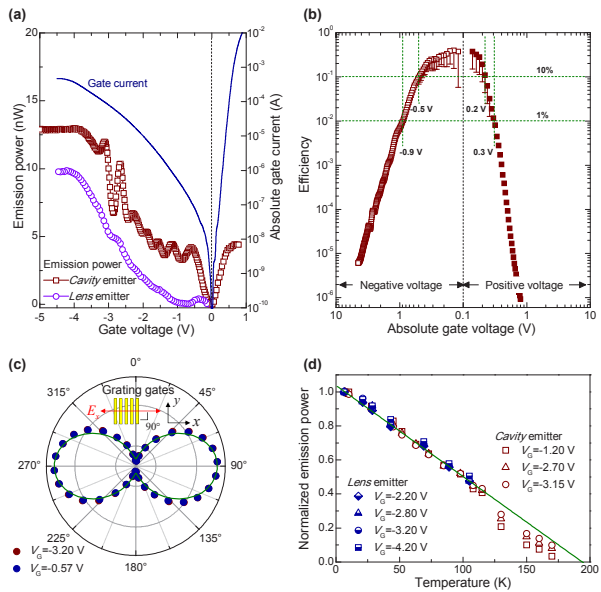}
\caption{
(a)~Gate current and emission power as a function of the gate voltage. 
(b)~Emission efficiency as a function of the absolute gate voltage. 
(c)~Polarization characteristic of the emission.
(d)~Temperature dependence of the emission power. 
}\label{fig_2}
\end{figure}

In both emitters at cryogenic temperature, non-equilibrium plasmons in the gated 2DEG are excited 
by a tunneling current between the gates and the 2DEG through the 25~nm thick GaN/AlGaN/AlN barrier.
The drain and source electrodes are connected to ground, as schematically shown in Fig.~\ref{fig_1}(b) and (c).
A negative gate voltage up to -4.5~V allows to reduce/deplete the electron density under the gates 
and inject electrons into the 2DEG simultaneously.
In contrast, a positive gate voltage up to +1.0~V allows for extraction of electrons from the 2DEG. 
Since the tunnel barrier is 25~nm thick, 
injection of tunneling electrons is rather weak and is assisted by the defect states in the barrier, as schematically shown in Fig.~\ref{fig_1}(f). 
The current-voltage characteristics of the emitters are shown in Fig.~\ref{fig_2}(a) 
revealing both Poole-Frenkel tunneling~\cite{poole-frenkel-1,poole-frenkel-2} 
with low field ($V_\mathrm{G}\gtrsim -2$~V) 
and Fowler-Nordheim tunneling with high field ($V_\mathrm{G}\lesssim -2$~V). 
Although the injection current density is of $10^{-7}-10^{-2}~\mathrm{A/cm^2}$,  
the emitted terahertz power less than 15~nW for both emitters can be well detected by using a silicon bolometer cooled at 4.2~K.  
We found that a direct terahertz field-effect transistor detector based on the same type of 2DEG cooled at 77~K 
is able to sense the terahertz emission as well (data not shown).
It has to be mentioned that 
defect-assisted Poole-Frenkel tunneling is significantly suppressed in those similar GaAs-based emitters 
since the crystal is nearly perfect and the tunnel barriers are thicker~\cite{hopfel,hirakawa,maranowski,otsuji,hirakawa93}.  

The wall-plug conversion efficiency defined as the ratio of the emitted terahertz power to the input electrical power 
is shown in Fig.~\ref{fig_2}(b).
The efficiency strongly depends on the gate voltage and thus the energy of the tunneling electrons. 
The efficiency reaches $10^{-2}$ and $10^{-1}$ with a negative gate voltage of -0.9~V and -0.5~V, respectively. 
With a more negative gate voltage than -1~V, the efficiency is reduced down to the order of $10^{-6}-10^{-3}$. 
Similarly, with a positive gate voltage so that electrons are extracted from the 2DEG, 
the efficiency decreases from $10^{-1}$ at +0.2~V to $10^{-2}$ at +0.3~V.  
In both cases, the lower the energy of the tunneling electrons, the higher the conversion efficiency. 
Furthermore, it is more effective to excite terahertz emission by injecting electrons into 
rather than extracting electrons from the 2DEG. 
Regulated by the grating, the emitted terahertz wave is linearly polarized along direction $x$, 
i.e., a TE mode, as shown in Fig.~\ref{fig_2}(c). 
At higher temperatures, increase in phonon scattering and reduction in electron mobility lower the emission efficiency, 
as shown in Fig.~\ref{fig_2}(d).
For both emitters, the emission power $P$ decreases linearly with temperature $T$.   
A linear fit based on $P=P_0(1-T/T_\mathrm{M})$ with $P_0$ the maximum achievable power 
yields a maximum operation temperature of $T_\mathrm{M} \approx 194$~K. 
When the temperature increases from 7~K to 77~K, about sixty percent of the emission power can be retained.

\begin{figure}[!htp]
\includegraphics[width=.4\textwidth]{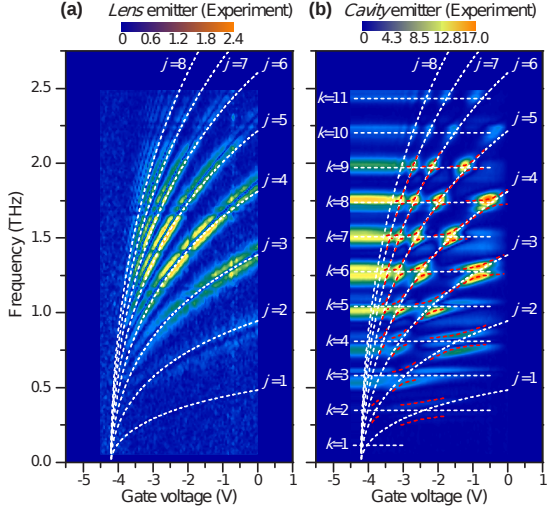}
\caption{
Measured emission spectra of the lens emitter (a) and the cavity emitter (b) at different gate voltages.
Dashed curves are calculated plasmon modes tuned by the gate voltage. 
Horizontal dashed lines are calculated F-P terahertz cavity mode frequencies. 
The red dashed curves are calculated plasmon-polariton modes. 
}\label{fig_3}
\end{figure}

The origin of the observed terahertz emission is revealed by measuring the emission spectra as a function of the gate voltage,  
as shown in Fig.~\ref{fig_3}(a) and (b). 
For the lens emitter, the continuously tunable terahertz emission spectrum from each discrete plasmon mode agrees well with 
the calculated plasmon modes (dashed curves) with mode number $j=1-8$. 
The emission from the first plasmon mode is hardly visible. 
For the cavity emitter, the emission spectra are remarkably different:  
terahertz emission occurs only when each plasmon mode becomes in resonance with one of the terahertz cavity modes. 
In Fig.~\ref{fig_3}(b), plasmon modes and F-P cavity modes are marked by the dashed curves and the horizontal dashed lines, respectively.
The locations of the resonances agree well with the calculation. 
Mode splitting resulted from the strong coupling between the plasmon modes and the terahertz cavity modes 
occurs at some of the resonances in Fig.~\ref{fig_3}(b). 
Hybridization of the terahertz cavity mode and the plasmon mode can be modeled as two-coupled oscillators and 
the resulting LPP ($\omega_{kj}^{-}$) and UPP ($\omega_{kj}^{+}$) modes can be expressed~\cite{kavokin}. 
\begin{widetext}
\begin{equation}
\omega_{kj}^\pm = \frac{\omega_k+\omega_{\mathrm{P}j}}{2} -\frac{i}{2}(\gamma_\mathrm{C}+\gamma_\mathrm{P}) \\
\pm \frac{1}{2}\sqrt{ (\delta\omega)^2+4V_{kj}^2 -(\gamma_\mathrm{C}-\gamma_\mathrm{P})^2-2i(\gamma_\mathrm{C}-\gamma_\mathrm{P})\delta\omega}, ~\label{Eq-1}
\end{equation}
\end{widetext}
where $\delta\omega=\omega_k-\omega_{\mathrm{P}j}$ 
is the detuning of the plasmon mode from the terahertz cavity mode, $V_{kj}$ is the coupling strength, 
$\gamma_\mathrm{P}$ is the linewidth of the plasmon mode, and $\gamma_\mathrm{C}$ is the linewidth of the terahertz cavity mode. 
Within the frequency range from 0.3 to 2.5~THz, the coupling strength can be divided into two regimes, 
namely the weak coupling regime with $k \geq 8$ and $j \geq 5$ and the strong coupling regime with $k \leq 7$ and $j \leq 6$. 
As shown in Fig.~\ref{fig_3}(b), 
calculation of the plasmon-polariton modes based on Eq.~(\ref{Eq-1}) agrees well with the observed mode splitting. 

\begin{figure}[!htp]
\includegraphics[width=.45\textwidth]{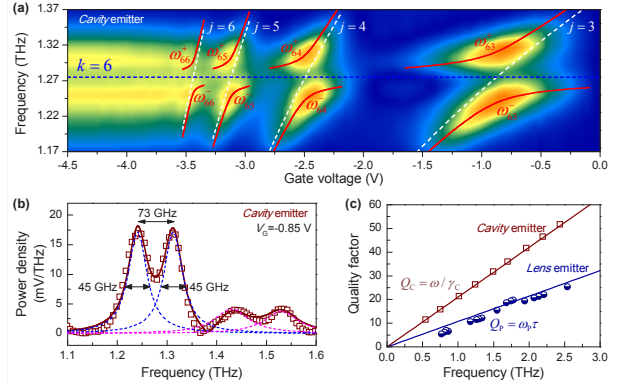}
\caption{
(a)~Zoom-in view of the plasmon-polariton modes formed from terahertz cavity mode $k=6$ and 
plasmon modes $j=3,4,5,6$. 
The solid curves are the corresponding LPP ($\omega_{kj}^{-}$)/UPP ($\omega_{kj}^{+}$) modes. 
(b)~Mode splitting with $k=6$ and $j=3$ at $V_\mathrm{G}=-0.85$~V. 
(c)~Quality factors for the F-P terahertz cavity and the plasmon modes. 
}\label{fig_4}
\end{figure}

A zoom-in view of the plasmon-polariton emission spectra from cavity mode with $k=6$ and 
plasmon modes with $j=3, 4, 5, 6$ are given in Fig.~\ref{fig_4}(a) 
with calculated plasmon-polariton modes ($\omega_{kj}^-$, $\omega_{kj}^+$) overlaid.  
A single trace at $V_\mathrm{G}=-0.85$~V, i.e., at the resonance of $\omega_6=\omega_{\mathrm{P}3} = 2\pi \times 1.275$~THz, 
is extracted and plotted in Fig.~\ref{fig_4}(b). 
Curve fits with multiple Lorentzian peaks agree well with the experiment spectrum. 
The linewidths of the terahertz cavity mode and the plasmon mode are found to be $\gamma_\mathrm{C}/2\pi=45$~GHz and 
$\gamma_\mathrm{P}/2\pi=115$~GHz, respectively.
The observed Rabi frequency of $\Omega_\mathrm{R}=\sqrt{4V_{63}^2-(\gamma_\mathrm{C}-\gamma_\mathrm{P})^2}=2\pi \times 73$~GHz 
yields a coupling strength of $V_{63}=2\pi\times45.1$~GHz. 
As shown in Fig.~\ref{fig_4}(c), 
the quality factor of the plasmon modes $Q=\omega_\mathrm{P}\tau$ is intrinsically limited by 
the momentum relaxation time $\tau^{-1}=e/m^*\mu \sim 120~\mathrm{GHz}$, 
the linewidth of cavity modes is limited by the quality factor of the terahertz cavity. 

Plasma in solid state suffers from high damping which has been the main hurdle for making realistic active plasmonic terahertz devices. 
In this work, we have shown that low-energy tunneling electrons can excite plasmons in 2DEG with an efficiency above $10^{-1}$. 
Such a high efficiency is a prerequisite for possible construction of high-power solid-state terahertz emitters. 
The excitation scheme relies on the energy relaxation of incident electrons via plasmon emission 
instead of using the previously proposed plasma instability method which involves a high acceleration electric field 
inevitably causing heating of the solid-state lattice. 
A resonant energy transfer/relaxation between the incident electrons and the plasmons would be the ideal approach to 
generate a large number of plasmons far from the equilibrium without heating up the host material. 
Plasmon-polariton states formed in strong coupling regime are ideal for 
efficient storage of both plasmons and terahertz photons and the efficient energy transfer in between. 
This would require a deliberately design including a mechanism allowing for injecting low-energy electrons with high current density, 
a high-quality plasmon cavity, a high-quality terahertz cavity and a strong coupling of both cavities.

In summary, we show that high efficiency, tunable and spontaneous terahertz plasmon emission and terahertz plasmon-polariton emission 
achieved in a grating-gate coupled 2DEG based on AlGaN/GaN heterostructure. 
The findings provoke in-depth search for high-quality terahertz plasmonic/electronic materials and 
invention of possible plasmon-based terahertz lasers. 
Furthermore, the studied plasmonic structure provides a platform for 
exploring the strong light-matter interaction in terahertz frequency regime. 

This work was supported by the National Basic Research Program of China (G2009CB929303), 
the National Key Research and Development Program of China (2016YFC0801203),  
the National Natural Science Foundation of China (60871077 and 61505242) and 
the Knowledge Innovation Program of the Chinese Academy of Sciences (Y0BAQ31001). 
The authors acknowledge support from the Nanofabrication Facility at 
Suzhou Institute of Nano-tech and Nano-bionics (SINANO).  
H.Q. thanks J.P. Kotthaus, R.A. Lewis and H. Yang for insightful discussions.



\end{document}